\documentstyle[preprint,aps,floats]{revtex}
\input psfig.tex
\begin{document}
\preprint{DAMTP-96-59, astro-ph/9606087}
\title{Sub-degree Scale Microwave Anisotropies from Cosmic Defects}
\author{Neil Turok}
\address{DAMTP, Silver St,\\
 Cambridge, CB3 9EW,  U.K.\\
Email: N.G.Turok@damtp.cam.ac.uk\\
}
\date{14/6/96}
\maketitle

\begin{abstract}
If current ideas about unified field theories are correct, macroscopic
cosmic defects may well exist. The observation of such an 
entity would have enormous significance for our understanding
of fundamental physics.
This paper points out a novel observable signature of cosmic texture
and global monopoles, 
namely strong hot spots in the cosmic microwave anisotropy
pattern on subdegree scales. This signal should be 
readily detectable by the next generation of anisotropy mapping
experiments. The signature arises from overdensities 
in the photon-baryon fluid generated by the gravitational attraction
of the defects. The angular power spectrum of the anisotropy fluctuations
on subdegree scales
is also calculated, for cosmic string, 
global monopoles, and texture.
\\
\end{abstract}

The prospect of
accurate maps of the 
temperature fluctuations on the cosmic microwave sky gives us 
hope of resolving one of nature's most profound questions, the origin of
large scale structure in the universe. The recently approved 
COBRAS-SAMBA and MAP satellites, the Very Small Array interferometer,
and other experiments, promise to provide cosmological 
data of unparallelled quality and quantity.

Cosmic defects have been of fascination to theorists ever since it
was realised that their presence was generic in unified field theories
of high energy physics \cite{kibble}. 
Indeed, if current ideas about the unification
of forces are correct, cosmic defects in one form or another 
are almost unavoidable. Two such defects occur in standard model physics -
namely electroweak textures and skyrmions, formed as a result of the
breakdown of the electroweak and strong chiral symmetries respectively.
Neither of these is relevant to the problem of structure formation,
but analogous defects formed through the breakdown of symmetries
at much higher, grand unification energy scales, do give rise to
interesting theories of structure formation in the universe
\cite{shvil}. Cosmic strings are the best known of these,
but global monopoles and textures are also promising candidates
\cite{ntswed}.
A period of inflation can, but does not necessarily, dilute the defects to
unobservably low densities. For example, recent work has 
has shown that they may be efficiently 
produced during the reheating process following inflation \cite{kls}.
The imminent maps of the microwave anisotropy offer a unique 
observational window on cosmic defects, and the possibility of 
discovering whether any of these relics of unification have survived
within our observable universe.

As mentioned, macroscopic cosmic defects 
provide  a possible mechanism for 
structure formation which is a valuable alternative
to the more popular scheme based on inflation.
The two theories could hardly
be more different - where the inflationary scheme is inherently
quantum mechanical, the cosmic defect mechanism is 
purely classical. The formation of fluctuations 
during inflation involves {\it linear} physics, whereas the defect
theories are highly {\it nonlinear}. Related to this, 
the inflationary fluctuations take the form of Gaussian
random noise, whereas the defect induced fluctuations are highly
non-Gaussian.

Recent work has emphasised further fundamental differences. First,
the defect theories are highly constrained by causality, so that
the fluctuations are completely uncorrelated beyond the horizon scale
at all times \cite{ct}, \cite{albrecht}, \cite{tcausal}, \cite{hsw}.
 And second, the nonlinearity of the theories means
that the evolution of each Fourier mode $\vec{k}$ in the perturbations 
is different, and cannot be predicted on the basis of the initial
conditions of that mode alone, because all the modes are coupled. 
This property is referred to as `incoherence' - modes of
the same wavenumber $k$ do not follow identical time 
evolution \cite{albrecht}. This has the effect of smearing the
`Doppler' peaks in the angular power spectrum of the anisotropies,
caused acoustic oscillations in the photon-baryon fluid just prior
to recombination.

Advances in computer power have recently made possible 
realistic calculations of microwave anisotropy maps in the defect theories.
Following the COBE satellite results, several groups
have reported calculations for the large angle cosmic microwave
anisotropy patterns \cite{br}, \cite{pst}, 
\cite{durrerzhou}, \cite{cs}. Unfortunately, 
the information available in a map with 7 degree resolution, 
such as the COBE 4 year maps, appears too limited to meaningfully
discriminate between the inflationary and  defect theories -
all these theories are compatible with the COBE measurements.

The first realistic computations of degree-scale CMB anisotropy 
from cosmic defects were reported
some time ago \cite{coulson}.
These were performed assuming a fully ionised 
universe, mainly for technical simplicity.
The present paper shall consider instead 
the case of standard recombination, in which much more
small scale structure is visible,
 on scales down to 10 arc minutes.
Below this scale, even with standard recombination,
photon diffusion 
on the last scattering surface begins to blur the anisotropy pattern.

At early times, redshifts $Z>>1100$, photons are tightly coupled 
to ionised matter, forming the photon-baryon-electron fluid. 
As the universe cools, the electrons and baryons undergo `recombination'
forming
neutral atoms, and the plasma quickly becomes transparent.
In the approximation that recombination happens instantaneously,
the microwave anisotropy seen
in a direction ${\bf n}$ on the sky is given by
\begin{equation}
\frac{\delta T}
{T}({\bf{n}})
 = \frac{1}{4}\delta_{\gamma}(i) - {\bf{n \cdot v_\gamma}}(i)
 - \frac{1}{2}\int_i^f d\tau \dot{h}_{ij} {\bf n}^i{\bf n}^j
\label{eq:s1}
\end{equation}
where $\tau$ is conformal time, $\delta_\gamma$ is the photon
fluid density contrast, and ${\bf v}_\gamma$ the velocity of the
photon-baryon fluid on the spherical shell representing
the `surface of last scattering'. 
The term ${1\over 4} \delta_\gamma$ is the 
intrinsic temperature contrast, and the second term is the
Doppler effect.
The last term involves the time derivative of the metric perturbation 
$h_{ij}$, and represents
the change in the proper path length 
due to a time-dependent gravitational field along the line
of sight. This is
the Sachs-Wolfe effect\cite{sw}. 

Equation (\ref{eq:s1}) is written in the synchronous gauge, 
sometimes criticised in the literature because it possesses
residual gauge freedom. However, this gauge is 
well suited to the cosmic defect theories discussed here. 
In these theories, 
the universe is assumed to be perfectly smooth before 
the symmetry breaking phase transition which produced the defects.
The perturbation  variables can therefore all be set to zero. 
This choice, called initially unperturbed 
synchronous gauge  has no residual gauge freedom.  
The nice property of this gauge is that the Einstein equations
are then completely causal, with the value of all perturbation variables
at a given spacetime point being completely determined by 
initial conditions within the point's backward light cone.
Gauges like the Newtonian gauge 
suffer from the problem that perturbations due
to anisotropic stresses, which are generally present in these theories,
propagate acausally \cite{vs}.

In initially unperturbed synchronous gauge, all the variables in
equation (\ref{eq:s1}) have an unambiguous physical meaning. 
The contrast $\delta_{\gamma}$ in the photon density 
is that observed by a local freely falling observer (for example
a cold dark matter particle). Likewise $\bf{v_\gamma}$ is the 
velocity of the photon-baryon fluid with respect to
cold dark matter. There is an additional Doppler term 
which may be attributed to the peculiar motion of the dark matter,
but that arises from an integration by parts 
of the last term \cite{coulson}.

The split represented in equation (\ref{eq:s1}) is useful
in a causal theory, because the first two terms are completely
determined by the source for perturbations within the backward sound cone,
with comoving radius
$\int_0^{\tau_{rec}}
 c_s(\tau) d\tau$, with $\tau_{rec}$ the conformal time at recombination
and $c_s< 1/\sqrt{3} $ the speed of sound in the photon-baryon fluid.
The last term depends on the perturbation source over a much
larger region of spacetime, filling the backward light cone of 
each point on the photon's trajectory. 
Apart from the earliest
contributions to the integral, in the vicinity of the 
last scattering surface, there is little correlation between
the first two
terms in (1) and the Sachs Wolfe integral. 
Furthermore, the latter produces an approximately scale invariant angular 
power spectrum, monotonically falling, and with no `Doppler' peaks 
(see ref. \cite{coulson}). It is therefore unlikely to be 
important in influencing the position  of the peaks in the 
power spectrum produced from the first two terms, and is 
a subdominant effect in the height of the peaks.
I have checked this explicitly 
in the texture model of 
reference \cite{ct}, where the approximation of ignoring the 
Sachs-Wolfe integral in (1) yields a primary peak in the angular power
spectrum higher
than the full result, but only by 20 per cent. 
Note that this approximation  
is certainly {\it not} good for a theory 
like inflation, where there are large correlations on 
superhorizon scales.

In this paper I report on calculations of the
first two terms alone, which dominate the anisotropy on  
scales from a degree down to several arc minutes.
Including the Sachs-Wolfe integral is certainly a goal for
future work, but the memory and CPU requirements 
would have severely compromised the 
resolution available.
The calculational technique involves an improved nonlinear sigma model
code in which local stress energy conservation
holds to an accuracy of a few per cent, and is quadratically
convergent in the timestep. This is coupled to 
a three dimensional leapfrog code for the dark matter (taken to be 
`cold' dark matter), neutrinos (treated as a fluid with viscosity),
and the photon-baryon-electron fluid, with a time-varying speed of sound.
For strings, the 
flat spacetime code of ref. \cite{coulson} is employed, with the modification 
that an extra fluid 
(with $c_s^2 = 1/3$) is 
introduced to account for the energy and momentum lost by
the strings due to a) the expansion of the universe, and b)
the emission of small loops and gravitational waves.
This ensures exact local conservation of the source 
stress energy tensor. Whilst the string network used is reasonably
realistic, having for example similar scaling density to
that reported in expanding universe codes, the string results 
have considerably larger systematic uncertainties than those for
texture and monopoles.

Simulations were performed for $256^3$ boxes, with the horizon
at last scattering representing 45 grid spacings. 
Tests of the source stress energy tensor indicate that the
fluctuating part achieves 
scaling form after as little as 5 
grid spacings, which is still significantly 
before matter-radiation 
equality. At the instant of recombination, 
slices across each box were taken at intervals of
32 grid spacings (no correlation was apparent between 
neighbouring slices), with the Doppler and intrinsic
terms stored separately. This produced 8 maps per simulation.
The maps are available via ftp
from the author.

The four colour Figures show representative maps for strings,
global monopoles, textures and standard cold dark matter.
The maps are 10$^o$ square, and have been smoothed 
with a Gaussian of FWHM 12'' (5 grid spacings),
to simulate the effect of
Silk damping. The maps show the temperature contrast $(\delta T/T)$
in units of one standard deviation, arising from the first two terms 
of equation (1).
In each case the cosmological parameters used were
$\Omega=1$, $\Omega_B=0.05$, $h=0.5$, $\Lambda=0$.
As mentioned above, because the Sachs Wolfe integral has been
omitted, these maps have slightly less power than they should on 
scales above one degree or so. But the 
most striking features in the maps occur on 
smaller scales, and these should be accurately represented. 

There are two ways in which the theories studied are 
manifestly different. First, the {\it scale} of the structure
present in the maps increases as one goes from strings to monopoles, 
 textures and standard inflation. This is a reflection of
the differences in angular power spectra, 
and in particular the shift in the primary peak in the power spectrum
(Figure 5).  
Second, the texture and monopole maps show clear `hot spots' 
in excess of 5 or 6 $\sigma$, which would be highly improbable
in a Gaussian theory. The fact that hot spots, and no cold spots 
are seen, is a simple result of the fact that the defects themselves
represent localised concentrations of energy, which {\it attract} 
the photon-baryon fluid, and cause a local rise in
the local density and temperature. 
Localised defects, like unwinding textures or annihilating 
monopole-antimonopole pairs,
do
so more effectively than extended objects like 
strings. After a hot spot is produced, I have established that
it is the presence of the somewhat diffused surviving 
defect stress energy in 
the region that prevents the photon density from oscillating negative
and producing a cold spot.

Let us now discuss quantitative measures of these differences.
The first is the angular power spectrum, defined as the
ensemble average 
$C_l= <|a_{lm}|^2>$, where the temperature on the sky is expanded
in spherical harmonics as $(\delta T/T) (\theta,\phi)
= \Sigma a_{lm} Y_{lm}(\theta,\phi)$.
In Figure 5, the quantity $l(l+1)C_l$, constant for a 
perfectly scale
invariant spectrum, is plotted against $l$. Multipole index $l$
is converted into an angular scale by 
noting that on small scales, the sky is nearly flat, and 
$l$ is effectively the two dimensional wavenumber. Modes
with index $l$ have a wavelength on the sky of $\Delta \theta =
(2 \pi /l )$.
Figure 5 shows the ensemble averaged power spectra 
for the four theories considered here. Each has been normalised 
to the COBE 4 year data, using the results of 
\cite{pst} for monopoles and texture, and \cite{coulson} for strings
\cite{gorski}.
The curves show the average of
3,4 and 6 runs (producing 8 maps each) 
for strings, monopoles and textures respectively. 
The fractional statistical error in the power at each $l$ is 
$\pm \sqrt{2/N}$ for $N$ uncorrelated maps - for the available ensembles 
this works 
out at approximately 20 per cent for textures, 25 per cent for monopoles,
30 per cent for strings. 
The dashed line
shows the result for the standard inflationary model.

The solid line shows the result for the texture model calculation
presented in \cite{ct}, when it is used to compute the same 
contributions to the anisotropy studied here. The $C_l$'s from the model 
have been multiplied by a factor 0.6 after COBE normalisation
(this correction is consistent with the estimated systematic errors
quoted in \cite{ct})
 - it appears that model overestimates the 
height of the primary peak by this factor. The position of 
the primary peak is however in excellent agreement with the predictions of
refs. \cite{ct}, \cite{durrergangui}.
Note that the approximation employed here
of neglecting the Sachs-Wolfe integral significantly underemphasises 
the second peak (at $l\sim 650$), but not the third,
which also appears visible in the full texture simulations.

The higher peaks do appear to be present in texture simulations 
at the locations expected from the model of \cite{ct}, but 
the statistics so far obtained are inadequate 
to fully resolve them. One should also worry about effects due to 
the non-scaling initial conditions, and other sources of noise
which would tend to smear the peaks. 
The techniques used here are not ideal for quantitatively determining 
the extent of the decoherence of the defect-induced perturbations
(nor indeed the $C_l$'s themselves), 
for which a Fourier space approach would be preferable.
Nevertheless,
secondary peaks are apparently visible in the intrinsic temperature
term alone (dashed lines). There are also hints of secondary peaks in the
cosmic string theory (the spacing between the spikes, 
$\Delta l \sim 280$ is what one expects for coherent 
acoustic oscillations - one has $\Delta(c_s k \tau_{rec}) \sim \pi$, 
and $\Delta l \sim \Delta (k\tau_0) \sim 50 \Delta( k\tau_{rec})$ ),
but the statistics
are again insufficient for any firm conclusion.
It is notable that the power spectra for individual maps
frequently show very clear peaks, with the appropriate spacing 
$\Delta l \sim 280$, but these are displaced from map to map in 
such a way that the ensemble average is smoother. This seems to 
be a clear signal of
decoherence \cite{albrecht}.

The calculations reported here, as mentioned above, significantly
underestimate the total power at lower values of  $l$. 
The scale invariant plateau produced by the Sachs-Wolfe integral
is at $l(l+1) C_l \sim 7 \times 10^{-10}$. For $300 <l<2000$
or so for textures and monopoles, and $400<l<2000$ for strings, 
the $C_l$'s shown include the dominant effects, 
but the Sachs-Wolfe integral may add 
$\sim 20-40 $ per cent to 
the power.
Above $l\sim 2000$, at least in the case of strings, 
the structure of the defect-induced Sachs-Wolfe signatures
produced after last scattering
start to contribute, leading to a tail of
high $l$ power which is unaffected by Silk damping \cite{batty}.
It is interesting to compare results with Bouchet et. al. \cite{bouchet},
who calculated only the Sache-Wolfe integral for strings, in 
a Minkowski-space approximation. They concluded that the contribution 
was roughly scale invariant, and 
the rms anisotropy from the interval $[Z,2Z]$ was $\sim 6 G \mu$,
with $G$ Newton's constant and $\mu$ the string tension.
Here I find the rms from the 
the intrinsic
and Doppler terms to be larger, $\sim 13 G \mu$. 
Thus the 
contributions studied here are 
likely to mask the linear signatures
of the strings, at least on angular scales above 10 arc minutes. 
These conclusions are consistent with  model calculations of the
string anisotropy power spectrum \cite{macf}.

The number of
maxima and minima of height $\nu \sigma$, where $\sigma$ is the 
standard deviation, provides a simple quantitative test
of the nonGaussianity of the maps. 
For a Gaussian theory, the number of peaks and minima 
per unit area, $dN/d \nu $
is straightforwardly calculable in terms of moments of the power
spectrum
\cite{be}. In this case the curves for maxima and minima are simply related
by the replacement  $\nu \rightarrow -\nu$.
Figure 6 shows the number of maxima and minima for the theories 
studied here. While the curves for {\it minima} take a form very similar
to what one expects for a Gaussian theory, the
curves for {\it maxima} show a clear excess of high peaks,
especially marked in the monopole and texture theories.
Note that this skewness was not present in the maps made by Coulson et. al.
\cite{coulson} of the Sachs-Wolfe contribution, 
it is entirely an effect of the `intrinsic temperature' and
`Doppler' terms studied here.

One can read off from
Figure 6
how much
sky coverage is needed in order to see one of these extreme events.
For example, there is approximately one peak in excess of $5 \sigma$ 
in the texture theory in each 300 square degrees, and in the 
monopole theory in each 100 square degrees.
By contrast, the inflationary theory predicts on average 
less than one such peak
on an entire sky.
The apparently skew shape of the texture and monopole maxima curves 
near the peak may also be a useful
discriminator.

In conclusion, the prospects for distinguishing between 
the inflationary and defect theories appear excellent. 
The power spectra of fluctuations are very different,
with the defect theories having the primary peak 
in the angular power spectrum shifted to smaller angular scales. 
And the nonGaussianity of the defect maps, with an
excess of high peaks on scales of order 20 arc minutes,
should be readily detectable by the 
VSA experiment or the COBRAS-SAMBA satellite. 

\centerline{\bf Acknowledgements}

I thank A. Albrecht, C. Barnes, R. Caldwell, 
G. Efstathiou, M. Hobson, J. Magueijo and
A. Lasenby, A. Sornberger and P.Shellard for useful comments. I thank 
R. Crittenden and P. Ferreira for collaboration in developing the 
string and fluid codes. I thank M. Hobson for supplying the
standard inflation map. This work was supported by grant number
AST960003P at the Pittsburgh Supercomputing Center, startup funds
from Cambridge University, 
grants from PPARC, UK, NSF contract
PHY90-21984, and the David and Lucile Packard Foundation.

\vfill\eject

\begin{figure}[htbp]
\centerline{\psfig{file=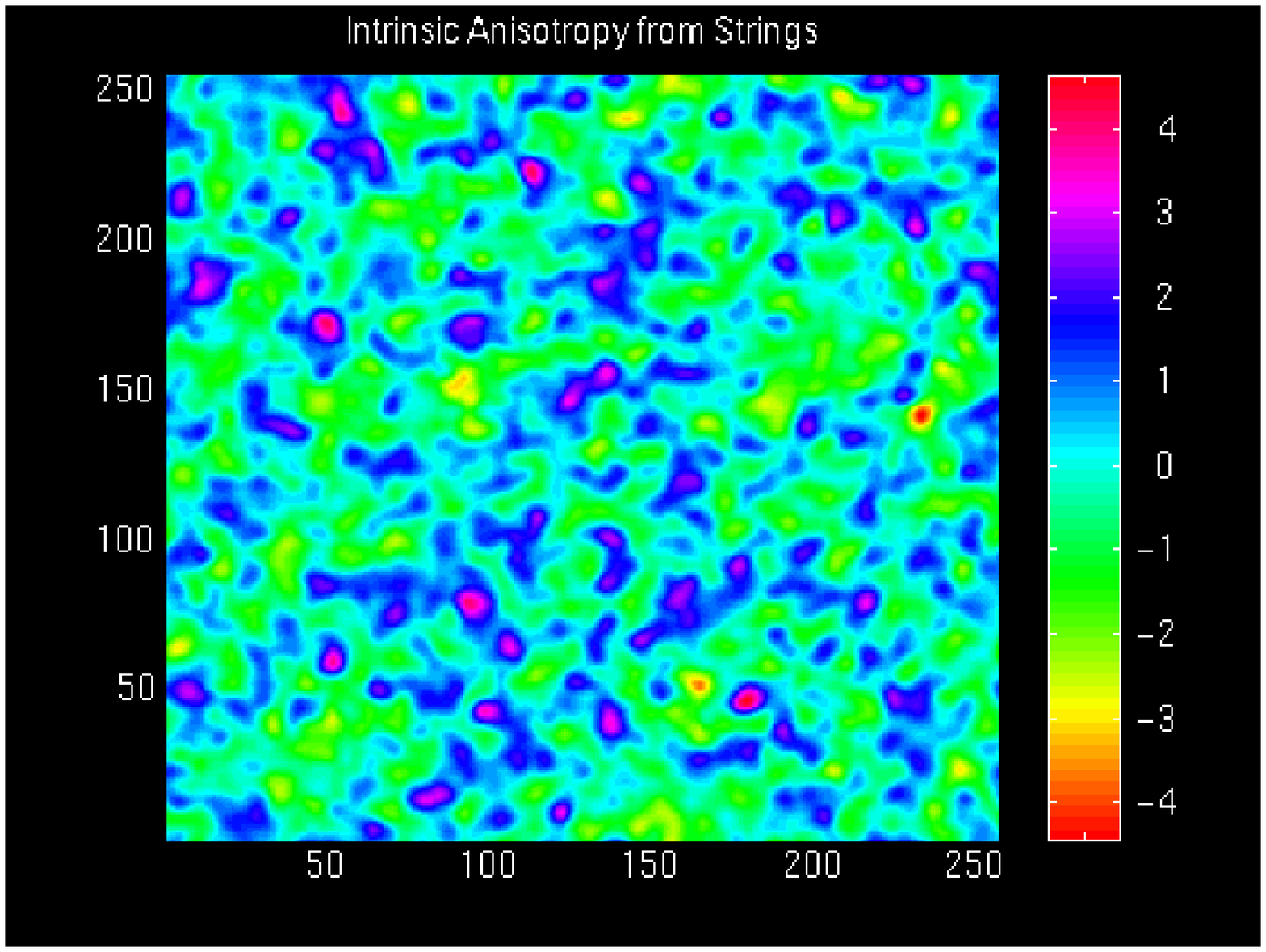,width=6.5in}}
\caption{ 
 A ten degree square patch of the microwave sky according to
the cosmic string theory for structure formation.
The colour scale shows the temperature contrast $\delta T/T$ in units of
one standard deviation. The numbers along the sides of the Figure
are grid units. 
The standard deviation for the string maps, 
adopting the COBE normalisation of ref. [14], but with a 
sky rms [18] on $10^o$ of 35 $\mu K$, 
is $\sigma= 2.5 \times 10^{-5}$.
}
\label{fig:f1}
\end{figure}
\vfill\eject

\begin{figure}[htbp]
\centerline{\psfig{file=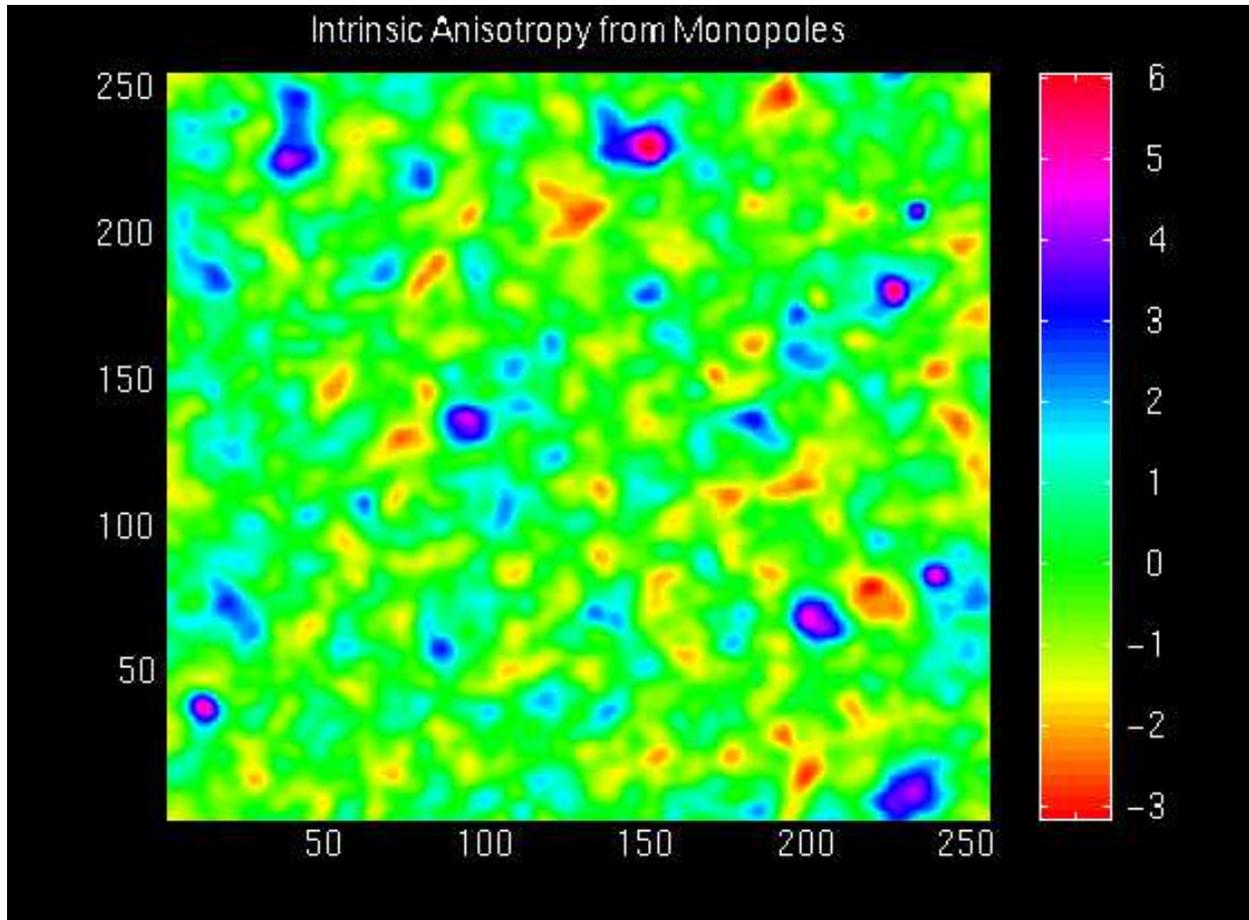,width=6.5in}}
\caption{As in Figure 1, but for global monopoles. 
The standard deviation for the monopole  maps, 
adopting the COBE normalisation of ref. [11], but with a 
sky rms [18] on $10^o$ of 35 $\mu K$, 
is $\sigma= 2.2 \times 10^{-5}$.
}
\label{fig:f2}
\end{figure}
\vfill\eject

\begin{figure}[htbp]
\centerline{\psfig{file=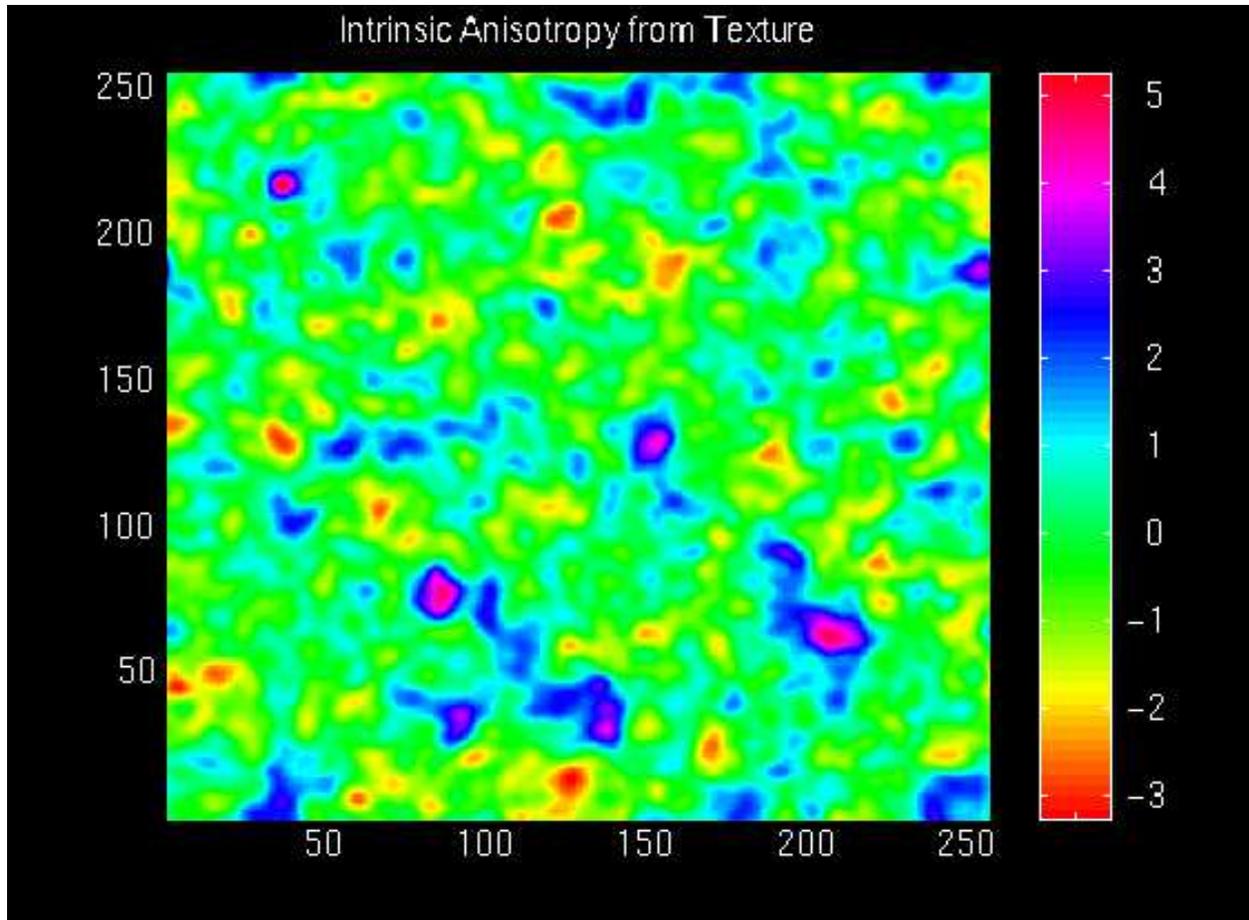,width=6.5in}}
\caption{ As in Figure 1, but for texture. 
The standard deviation for the texture  maps, 
adopting the COBE normalisation of ref. [11], but with a 
sky rms [18] on $10^o$ of 35 $\mu K$, 
is $\sigma= 2.3 \times 10^{-5}$.
}
\label{fig:f3}
\end{figure}
\vfill\eject

\begin{figure}[htbp]
\centerline{\psfig{file=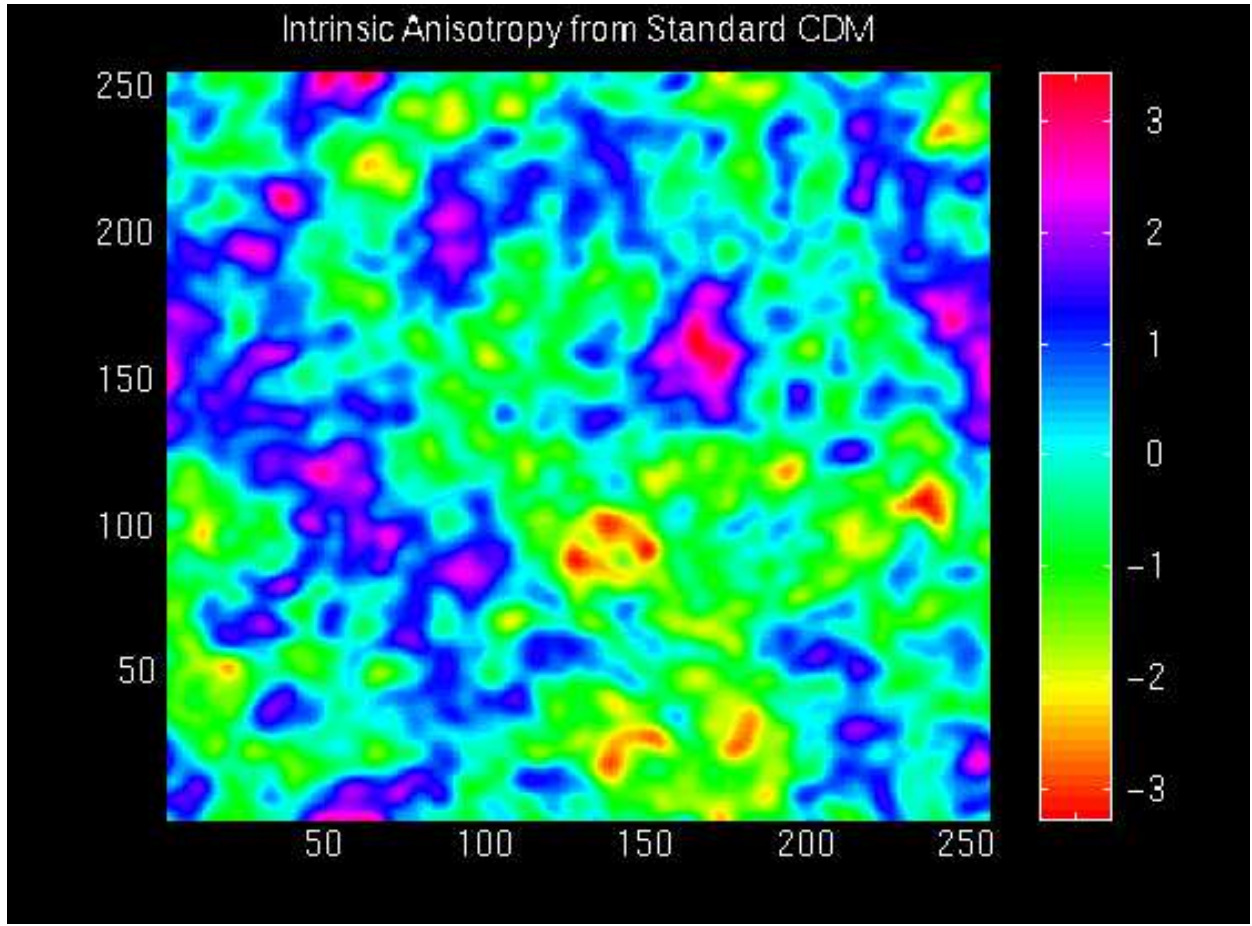,width=6.5in}}
\caption{ As in Figure 1, but for standard inflation.
}
\label{fig:f4}
\end{figure}
\vfill\eject

\begin{figure}[htbp]
\centerline{\psfig{file=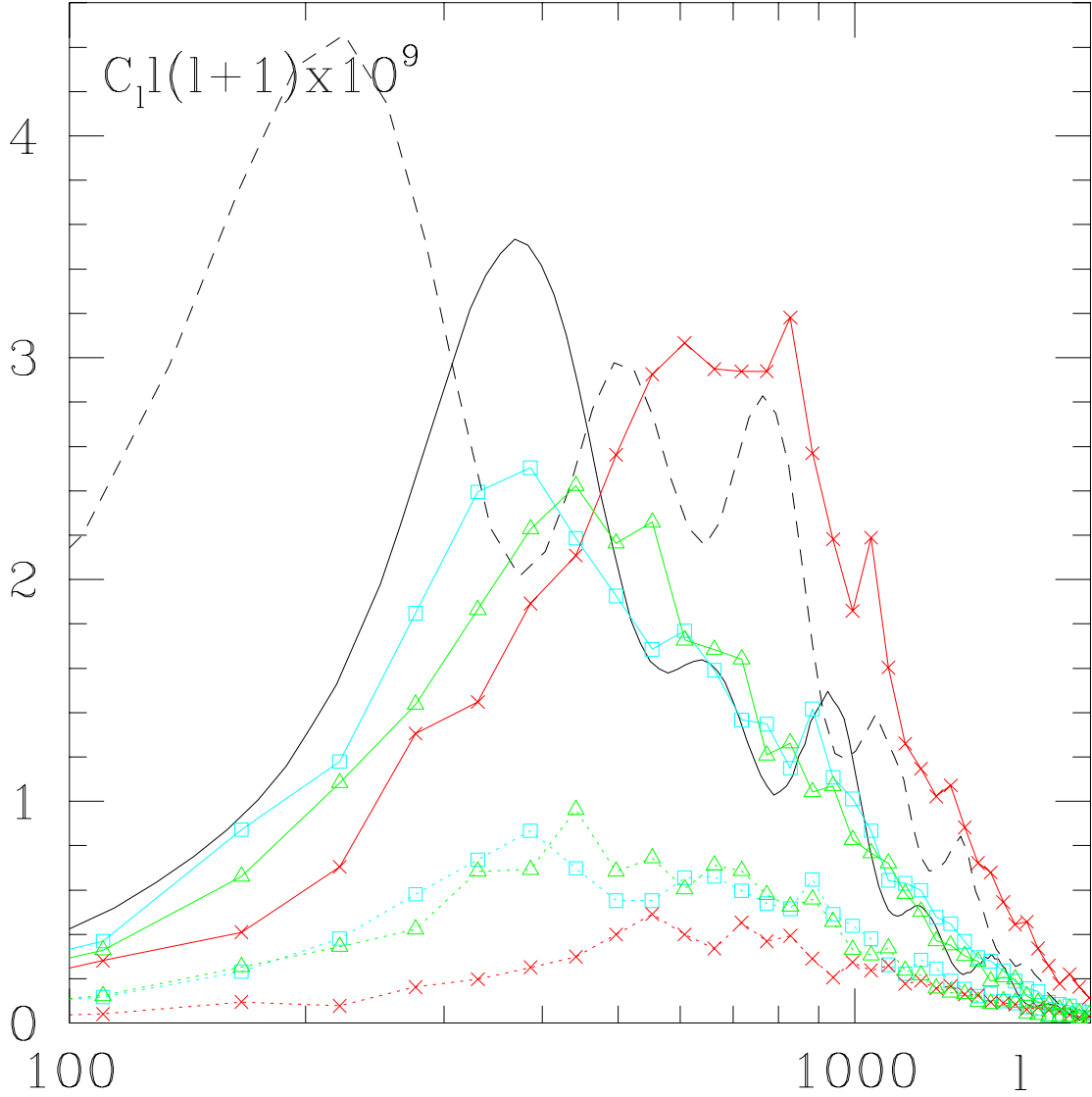,width=6.5in}}
\caption{ The angular power spectrum of intrinsic CMB anisotropies 
in the cosmic defect theories. The anisotropy is 
expanded in spherical harmonics, $(\delta T/T) = \Sigma a_{lm}
Y_{lm}(\theta,\phi)$, and $C_l$ is
defined as $<|a_{lm}|^2>$. 
The solid red, green and blue lines, linking crosses, triangles and 
boxes respectively, show the numerically determined power
spectra 
for cosmic strings, 
global monopoles and textures, when the theories are 
normalised to COBE according to 
the results of refs. [11] and [14]. The dashed lines show the
contribution from the intrinsic temperature term $(\delta T/T)_i$
alone (see text). The solid black curve shows the results
for the texture model of reference [5], scaled down by a factor 
of 0.6 after normalising to COBE at low $l$.
The 
dashed line shows the full $C_l$ spectrum for the 
`standard' 
inflationary theory.  
}
\label{fig:f5}
\end{figure}
\vfill\eject
 
\begin{figure}[htbp]
\centerline{\psfig{file=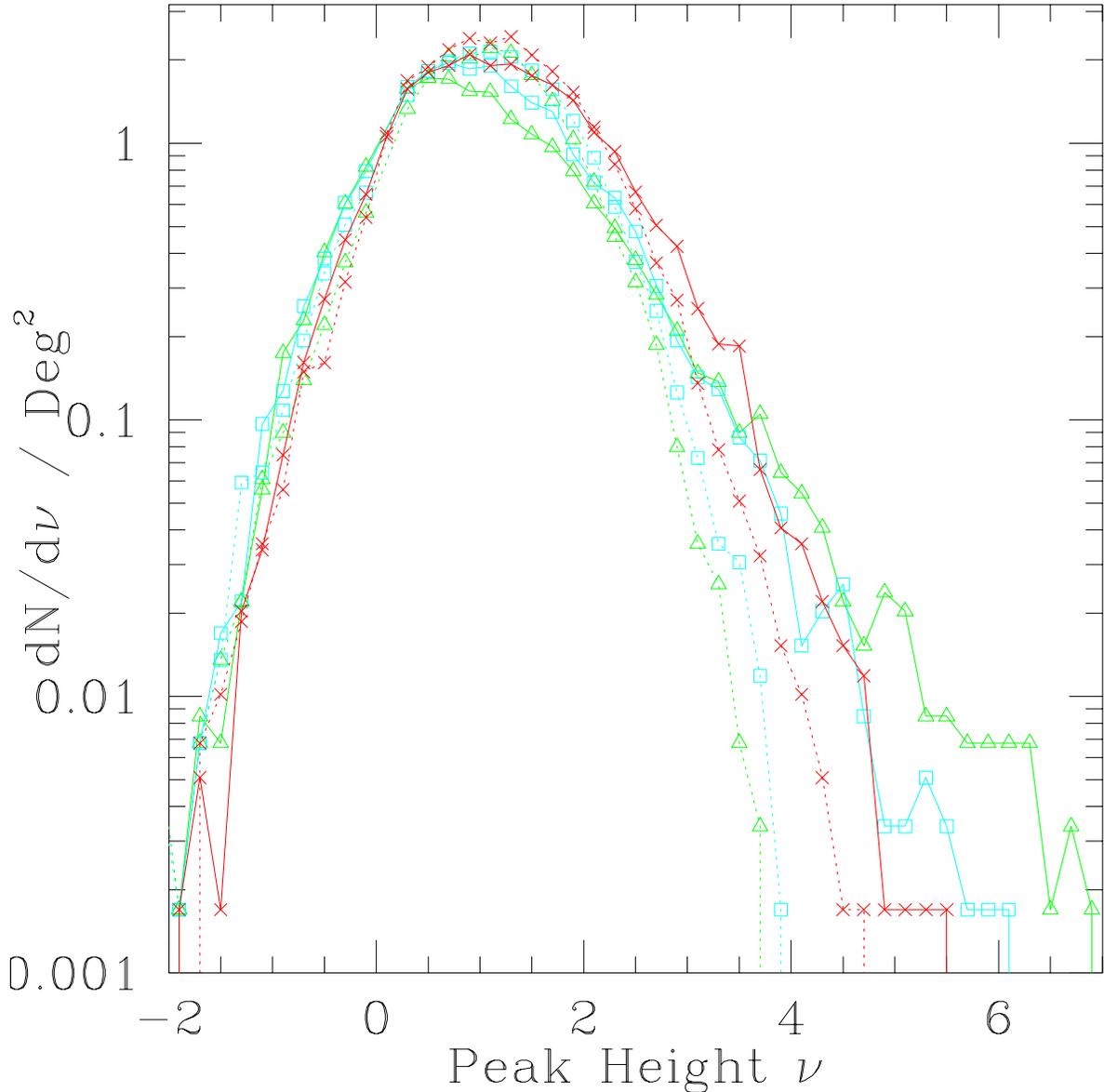,width=6.5in}}
\caption{ The differential number of maxima and minima of the temperature maps
in a given range of $(\delta T/ T)= \nu \sigma$, measured in units of 
the standard deviation $\sigma$. The solid red, green and 
blue lines, linking crosses, triangles and boxes respectively, show 
the results for maxima for cosmic strings, monopoles and textures. 
The dashed lines show the number of minima plotted against the 
{\it negative} of the height 
in the same theories. In a Gaussian theory, the curves for
maxima and minima are identical, and are computable from the
power spectra $C_l$ [21]. The curves shown for minima 
resemble those for a Gaussian random field, whilst the curves for maxima
are clearly nonGaussian, most markedly so for monopoles and textures.
}
\label{fig:f6}
\end{figure}
\vfill\eject

\end{document}